\documentclass[12pt]{article}
\usepackage{epsfig}
        \newdimen\eqskip
        \newdimen\txtskip
        \baselineskip=\txtskip
        \newdimen\mysep                
        \newdimen\hmysep
\begin{document}
  \newcommand{\ccaption}[2]{
    \begin{center}
    \parbox{0.85\textwidth}{
      \caption[#1]{\small{{#2}}}
      }
    \end{center}
    }
\newcommand{\BS}{\bigskip}
% MATH SYMBOLS
\def    \be             {\begin{equation}}
\def    \ee             {\end{equation}}
\def    \beq             {\begin{equation}}
\def    \eeq             {\end{equation}}
\def    \ba             {\begin{eqnarray}}
\def    \ea             {\end{eqnarray}}
\def    \beqn           {\begin{eqnarray}}
\def    \eeqn           {\end{eqnarray}}
\def    \beeq           {\begin{eqnarray}}
\def    \eeeq           {\end{eqnarray}}
\def    \nn             {\nonumber}
\def    \=              {\;=\;}
\def    \frac           #1#2{{#1 \over #2}}
\def    \ret            {\\[\eqskip]}
\def    \ie             {{\em i.e.\/} }
\def    \eg             {{\em e.g.\/} }
\def \lsim{\mathrel{\vcenter
     {\hbox{$<$}\nointerlineskip\hbox{$\sim$}}}}
\def \gsim{\mathrel{\vcenter
     {\hbox{$>$}\nointerlineskip\hbox{$\sim$}}}}
\def    \bentarrow      {\:\raisebox{1.1ex}{\rlap{$\vert$}}\!\rightarrow}
\def    \rd             {{\mathrm d}}    
\def    \Im             {{\mathrm{Im}}}  
\def    \bra#1          {\mbox{$\langle #1 |$}}
\def    \ket#1          {\mbox{$| #1 \rangle$}}
\def    \to             {\rightarrow} 

% UNITS                 
\def    \kev            {\mbox{$\mathrm{keV}$}}
\def    \mev            {\mbox{$\mathrm{MeV}$}}
\def    \gev            {\mbox{$\mathrm{GeV}$}}

% KINEMATICAL VARIABLES 

\def    \mq             {\mbox{$m_Q$}}  
\def    \mt             {\mbox{$m_t$}}  
\def    \mb             {\mbox{$m_b$}}  
\def    \mqq            {\mbox{$m_{Q\bar Q}$}}
\def    \mqqsq          {\mbox{$m^2_{Q\bar Q}$}}
\def    \pt             {\mbox{$p_T$}}
\def    \ptsq           {\mbox{$p^2_T$}}

% QCD PARAMETERS                                      
\newcommand     \MSB            {\ifmmode {\overline{\rm MS}} \else 
                                 $\overline{\rm MS}$  \fi}
\def    \muf            {\mbox{$\mu_{\rm F}$}}
\def    \mug            {\mbox{$\mu_\gamma$}}
\def    \mufsq          {\mbox{$\mu^2_{\rm F}$}}
\def    \mur            {{\mbox{$\mu_{\rm R}$}}}
\def    \mursq          {\mbox{$\mu^2_{\rm R}$}}
\def    \mul            {{\mu_\Lambda}}
\def    \mulsq          {\mbox{$\mu^2_\Lambda$}}

\def    \bzero          {\mbox{$b_0$}}
\def    \as             {\ifmmode \alpha_s \else $\alpha_s$ \fi}
\def    \asb            {\mbox{$\alpha_s^{(b)}$}}
\def    \assq           {\mbox{$\alpha_s^2$}}
\def \oacube {\mbox{$ O(\alpha_s^3)$}}
\def \oafour {\mbox{$ O(\alpha_s^4)$}}
\def \oatwo {\mbox{$ O(\alpha_s^2)$}}
\def \oas   {\mbox{$ O(\alpha_s)$}}
\def\asp{{\alpha_s}\over{\pi}}

\def\slash#1{{#1\!\!\!/}}
\def\rt1{\raisebox{-1ex}{\rlap{$\; \rho \to 1 \;\;$}}
\raisebox{.4ex}{$\;\; \;\;\simeq \;\;\;\;$}}
\def\ltap{\raisebox{-.5ex}{\rlap{$\,\sim\,$}} \raisebox{.5ex}{$\,<\,$}}
\def\gtap{\raisebox{-.5ex}{\rlap{$\,\sim\,$}} \raisebox{.5ex}{$\,>\,$}} 

\def\GE{\gamma_E}
\def\half{\frac{1}{2}}
\def\b0{\beta_0}
\def\naive{na\"{\i}ve}
\def\cm{{\cal M}}
\def\bom#1{\mbox{\bf{#1}}}
%%%%%%%%%%%%%%%%%%%%%%%%%%%%%%%%%%%%%%%%%%%%%%%%%%%%%%%%%%%%%%%%%%%%%%
\begin{titlepage}
\nopagebreak       {\flushright{
        \begin{minipage}{5cm}
        CERN-TH/98-249\\
        {\tt hep-ph/9807570}\\
        \end{minipage}        }

}
\vfill
\begin{center}
{\LARGE { \bf \sc A new approach to multi-jet \\[0.5cm]
calculations in hadron collisions~\footnote{This work was supported in part 
by the EU Fourth Framework Programme ``Training and Mobility of Researchers'', 
Network ``Quantum Chromodynamics and the Deep Structure of
Elementary Particles'', contract FMRX--CT98--0194 (DG 12 -- MIHT).}}}
\vfill                                                       
{\sc      F. Caravaglios$^a$, 
          M.L. Mangano$^a$  
    \footnote{On leave of absence from INFN, Sezione di Pisa, Italy.},
          M. Moretti$^{a,b}$
         and R. Pittau$^a$
}
\\[1cm]                  
{$^a$ CERN, Theoretical Physics Division, \\ CH~1211 Geneva 23, Switzerland} 
\\[0.5cm]                  
{$^b$ Dept. of Physics and INFN, Ferrara, Italy}
\end{center}                             
\nopagebreak
\vfill
%\vskip 3cm
\begin{abstract} 
We present an algorithm to evaluate the exact, tree-level matrix elements 
for multi-parton processes in QCD. We tested this
technique, based on the recursive evaluation of the $S$-matrix, 
on  processes such as $gg \to n$ gluons and $q\bar q \to n$ gluons,
with $n$ up to 9. The summation over colour configurations is designed to allow
the construction of parton-level event generators suitable to interfacing with
a parton-shower evolution including the effects of colour-coherence. This leads
to a fully exclusive, hadron-level description of multi-jet final states,
accurately incorporating the dynamics of large jet-jet separation angles.
Explicit results for the total rates and differential distributions of
processes with 8 final-state partons are given.
\end{abstract}                                                          
\vskip 1cm
CERN-TH/98-249\hfill \\
August 1998     
\vfill       
\end{titlepage}

\section{Introduction}
\label{introduction}
Multi-jet final states play an important role in the study of high-energy
collisions. They provide in fact interesting signatures for several phenomena,
both within the Standard Model (e.g. top-pair production), and beyond it
(e.g. multi-jet decays of supersymmetric particles such as gluinos and
squarks).  The accurate determination of the properties of these phenomena
requires a good understanding of the properties of the usually large
multi-jet QCD backgrounds, which can distort the shapes of signal distributions
and affect the measurement of quantities such as resonances' masses.
In the past few years, several theoretical developments have allowed the
calculation of very complicated multi-parton processes in
QCD (for a review, see~\cite{review})~\footnote{We shall include under the
label QCD also all processes with the emission of electro-weak gauge bosons,
such as associated production of $W$'s and jets.}.
For example, the complete set of leading-order (LO) background processes to the
production and decay of top-quark pairs in hadronic collisions is known for
both the fully hadronic decays~\cite{Berends89b}, and the $e\nu$+4-jet
decays~\cite{vecbos}.                                          

While significant progress has been made in the field of one-loop
corrections (for a review see~\cite{Bern92}), quantitative studies of multi-jet
processes (with $n > 3$) can only be done today using tree-level results,
which are the subject of the present work.      
There are several reasons for wanting to improve the tools currently available
to perform these calculations.           
\begin{enumerate}                             
\item First of all, interesting final states with larger jet multiplicities
will become available with the next generation of colliders (LHC and       
NLC). This will hold both for standard QCD processes, where the huge available
phase-space will allow production of many high-$E_T$ jets, and for potential
signals of new physics (a good example being cascade decays of heavy squarks or
gluinos with $R$-parity non-conservation, where final states with over 10 jets 
would be a typical signature).                      
This calls  for  improved algorithms, to allow the calculation of cross-sections
for complicated processes within reasonable amounts of computer time.
\item Secondly, one would like to be able to complement the calculation of
parton-level matrix elements with the evaluation of the full hadronic structure
of the final state. This is a key ingredient 
for a satisfactory study of both signal and background components and
for a complete comparison between theory and data. It involves the consistent
merging of the matrix-element computation with the parton-shower evolution, a
problem which has not been considered in the development of previous tools for
the evaluation of multi-jet processes.
\end{enumerate}

While relations are known~\cite{Berends88} which allow to systematically
evaluate high-order tree-level processes in a recursive fashion, the complexity
of the algorithm grows very quickly and makes progress beyond the processes
listed above very hard. Several approximations have therefore been
introduced~\cite{sphel,Maxwell87}, to evaluate with rather low computing time
and acceptable accuracy cross-sections for many partons in the final state.
Nevertheless, the knowledge of the exact parton level matrix elements will
always be needed, in order to assess the reliability of the approximations
used.
                                                  
In this work we present an approach to the calculation of tree-level matrix
elements for multi-parton final states which addresses both of the above
problems, and therefore improves on the currently available algorithms.  The
key element of this proposal  is the use of the algorithm {\tt ALPHA},
developed by two of us~\cite{alpha} for the evaluation of arbitrary
multi-parton matrix elements. This algorithm determines the matrix elements
from a (numerical) Legendre transform of the effective action, using a
recursive procedure which does not make explicit use of Feynman diagrams. 
The algorithm has a complexity growing like a
power in the number of particles, compared to the factorial-like growth that
one expects from naive diagram counting.
This is a necessary feature of any attempt to evaluate matrix elements for
processes with large numbers of external particles, since the number of Feynman
diagrams grows very quickly beyond any reasonable value.
For example, the number of tree-level Feynamn
diagrams corresponding to a process with $n_g$ gluons and $n_q$ pairs of quarks
(with $n_q=0,1$), 
is given by the following formula
\footnote{
The formula can be easily derived as follows. For a given diagram consider the
triplet $(p,q,g)$, where 
$p$ is the number of 3-gluon vertices, $q$ is the total number
of external quark legs and
internal quark propagators,  and $g$ is the total number of
gluonic external legs and internal propagators. The number of diagrams obtained
by attaching an additional gluon in all allowed ways is given by:
$p$ diagrams of type $(p-1,q,g+1)$, plus $q$ diagrams of type $(p,q+1,g+1)$
plus $g$ diagrams of type $(p+1,q,g+2)$.
These triplets can be obtained by applying the 
operator $y \partial / \partial x + x y^3 \partial / \partial
y + z^2 y \partial / \partial z $ to the function $x^p \, y^g \, z^{q}$. 
Their multiplicities are given by the coefficients of the relative monomials,
and the                                                                      
total number of generated diagrams is extracted by setting $x=y=z=1$ in the
polynomial thus obtained.                       
Repeated iteration of this operator on the three-point diagrams $ggg$ and
$q\bar q g$, represented by the monomials $x\, y^3$ and $y \, z^2$, gives the
desired result.}:
\be \label{oper}
N_{diag} \=
\left[ \left(y {\partial \over \partial x}+ x y^3 {\partial \over \partial
y} + z^2 y {\partial \over \partial z} \right)^{n_g+2n_q-3} 
 \; x^{1-n_q} \, y^{3-2n_q} \, z^{2n_q} \right]_{x=y=z=1}
\ee                                                
A similar formula can be constructed for processes with more than 1 quark pair.
The number of diagrams relevant for some of the examples considered in this
paper are collected in Table~1. 
\begin{table}          
\begin{center}
\begin{tabular}{|c|r|r|r|r|} \hline 
 \rule[-7 pt]{0 pt}{24 pt} 
Process                 & $n=7$ & $n=8$ & $n=9$ & $n=10$ \\
\hline                                                     
\hline
\rule[-7 pt]{0 pt}{24 pt}
$ g~g      \to n\; g$   & 559,405 & 10,525,900 & 224,449,225 & 5,348,843,500 \\
\hline                                                               
\rule[-7 pt]{0 pt}{24 pt}
$ q \bar q \to n\; g$   & 231,280 & 4,016,775 & 79,603,720 &  1,773,172,275 \\
\hline                                                             
\end{tabular}
\ccaption{}{Number of Feynman diagrams corresponding to amplitudes with
different numbers of quarks and gluons.}
\end{center}
\end{table}
These numbers clearly illustrate the problems encountered when trying to
evaluate the amplitudes by calculating each individual diagram, whether by
algebraic or by numerical means.                                        

{\tt ALPHA} has been shown to operate very successfully in the case of
purely electroweak processes~\cite{alpha-appl}, and will be reviewed here in
Section~2. Its application to the case of QCD, although straightforward,
requires  some care  due to the rapid increase of the number of colour
configurations with the number of external coloured particles.  Furthermore,
the choice of how to organise the sum over colour structures has important
implications for the possibility to merge the evaluation of a given matrix
element with the successive parton-shower QCD evolution. Several options are
available, in principle, to deal with the summation over all possible colour
configurations. They will be discussed in Section~3, where our strategy for an
efficient generation of Monte-Carlo events, combined with the possibility to
generate colour configurations suitable for a parton-shower evolution, will be
presented.
Some numerical examples of applications of this tecnique to the case of 
$2\to 8$ parton processes in hadronic collisions are given in Section~4.
A more complete study, including the treatment of associated production of jets
and electroweak gauge bosons 
and the description of a complete event generator
interfaced to a parton-shower program such as {\tt HERWIG}~\cite{herwig}, will be
presented in the future.

Independently of our efforts, work by Draggiotis, Kleiss and
Papadopoulos~\cite{Kleiss98} has recently addressed the problem of the efficient
generation of multi-parton QCD final states using the {\tt ALPHA} algorithm. In
this work they turn the summation over colours into an integration over a
continuous set of variables, slightly gaining in computational accuracy over
the more standard colour-summation technique presented in our work. The
integration technique, however, does not lend itself in an obvious way to the
efficient merging of the parton-level calculation with the parton-shower
evolution.        In either case, it is quite clear that improvements in the
numerical efficiency of these calculations will be possible, and work should be
devoted in the future to find the best compromise among all different
requirements to be met for a faithful and efficient  representation of these
highly complex processes.
                                                 
\section{The {\tt ALPHA} algorithm}
In reference \cite{alpha}, a new approach to the computation 
of tree level scattering amplitudes was introduced.
This approach, based on the numerical  Legendre transform of the effective 
action, is  particularly useful for the automatic calculation of multi-particle 
processes. 
This technique was implemented in a {\tt Fortran} code 
\cite{alpha} which has                                       
been succesfully used  to study several intricated electroweak processes 
\cite{alpha-appl}. For the sake of completeness, we review in this Section 
the {\tt ALPHA} algorithm; the interested reader can find                  
a more detailed discussion in the 
original  paper \cite{alpha}, which includes
an explicit analytic example for the $\lambda \phi^3 $ theory.
                     
Let $\Gamma$ be the one-particle-irreducible generator of the 
Green functions for a given theory. Then 
the computation of the S-matrix requires the evaluation 
of  the  Legendre transform, Z, of $\Gamma$:
\be                                                      
Z(J^\alpha)= - \Gamma(\phi^\alpha) 
+ J^\alpha(x)\phi^\alpha(x)     
\label {legtr}
\ee
where $\phi^\alpha$ are the classical fields defined as the solutions of
\be                                                                  
 J^\alpha =\frac  {\delta \Gamma }{\delta \phi^\alpha} \; ,
\label{minim}                                              
\ee
and the $J^\alpha$  
play the role of classical sources.
 In general the Lagrangian contains several fields, with different spin 
and internal quantum numbers. It may also  contain interactions with an 
arbitrary number of fields; for our purposes it is better to rearrange the
 Lagrangian into an equivalent form  which includes only trilinear interactions:
this can be achieved by  introducing a proper set of auxiliary fields\footnote{
 For example a term of the type $\lambda \phi^4 $ can be replaced by the 
equivalent form 
$ \lambda \phi^2 X -1/4  X^2$, where $X$ is an auxiliary field. The equation 
of motion for the auxiliary field, $X=2\lambda \phi^2$,  makes this
 equivalence manifest.}. 
The  Lagrangian in momentum space is   
\ba \label{lagrr}
{\cal{L}}&=&{1\over 2 } \int{{d^4 \bar p}~ {d^4 \bar q }~
\bar\delta(p+q)~ 
\tilde\Pi^{\alpha\beta}(p^2)~ \phi^\alpha(p)~ \phi^\beta(q)}+\nn\\
&+&{1\over 6} \int{{d^4 \bar p}~ {d^4 \bar q }~
{d^4 \bar k }~
 \bar \delta(p+q+k)~ \phi^\alpha(p) ~\phi^\beta(q)~\phi^\gamma(k)~
 {\cal{O}}^{\alpha\beta\gamma}(p,q,k)} \; , \nn
\ea                                         
where the greek indices are a compact notation for Lorentz indices,
internal symmetry indices, flavour etc. and where $d^4 \bar p=d^4 p/ (2 \pi)^4 $
and $\bar \delta=(2 \pi)^4 \delta^4$.                 
$\tilde\Pi^{\alpha\beta}(p^2) $ is the inverse propagator and   
$ {\cal{O}}^{\alpha\beta\gamma}(p,q,k)$ is a generic function of the momenta.
In the  case of translationally invariant local interactions ${\cal{O}}$ is  a
polynomial in the momenta.
% (in our compact notation $\alpha,\beta,\gamma$ 
%contain possible Lorentz indices to be contracted with the spin indices  of the
%fields  $\phi^\alpha(p)$ etc.).   
To obtain the connected Green function 
$G^{\alpha\beta\cdots\gamma}(p_1 \cdots p_n)$ from the above Lagrangian 
 we 
introduce the classical sources 
\be
J^\alpha(q)=\sum_{i=1}^n a_i^\alpha \bar \delta(q-p_i)
\label{source}
\ee
where the $a_i^\alpha$  carry  the same quantum
numbers as the source  $J^\alpha$.
With this choice of the  $J^\alpha$ the amplitude $\cal A$ is given
by\footnote{For the sake of simplicity, we omit from this expression the
explicit truncation of the external propagators.}:
\be \label{deriva}                                
{\cal A}=\frac{\partial Z}{\partial a_1^\alpha \dots \partial
 a_n^\gamma}|_{a_1^\alpha=0,\dots,a_n^\gamma=0} 
\ee
The equations of motion of eq.~(\ref{minim})  can be written  as 
\be                  \label{eqmoto}           
\phi^\alpha(q)=\tilde \Pi_{\alpha\lambda}^{-1}(q)\left[ J^\lambda(q) - 
{1\over 2} \int{{d^4 \bar p}}~
{d^4 \bar k }~
\bar \delta(q+k+p)~                   \phi^\beta(k)~\phi^\gamma(p)~
 {\cal{O}}^{\lambda\beta\gamma}(q,k,p)\right].
\ee
It is clear that this equation for $\phi(q) $  can be solved perturbatively 
with respect to the interaction $ {\cal{O}}^{\alpha\beta\gamma}$
 (or, equivalently, with respect to the $a_i^\alpha$) and the $(t+1)$-th  
order  of this perturbative series is obtained by inserting the expansion of the 
 $\phi^\alpha(q)$ up to the $t$-th order in the right-hand side of  
eq.~(\ref{eqmoto}). 
To recover the functional derivative of eq.~(\ref{deriva}) and avoid 
unnecessary 
computations it is useful to introduce the prescription to drop out terms which 
 contain powers of the $a_j^\alpha$ larger than one, at  any iteration step.
With this prescription, 
using the initial condition (\ref{source})  and the recursive relation 
(\ref{eqmoto}),
we can  prove by induction that the solution $\phi^\alpha(q)$ is of the form 
\be \label{eqphi}                                                
 \phi^\alpha(q)=\sum_{j=1}^{2^n-2} b_j^\alpha~ \bar \delta(q-P_j)
\ee                                                       
with 
\be\label{choice}
P_j=c^i_j~ p_i~~~~~~~~~~c^i_j=0,1.
\ee
%Each  $b_j^\alpha$ can be computed exploiting the \ref{eqmoto} after the
% substitution \ref{eqphi}; if a $b_j^\alpha$ corresponds to a momentum 
%$P_j=c^i_j p_i$ with a $c^i_j>1$, then one can prove by induction applying  the
% \ref{eqmoto} that  $b_j^\alpha \sim (a^\alpha)^{c^i_j}$ and will vanish in 
%the limit $a^\alpha \rightarrow 0$ needed to extract the Green function from 
%the generating functional $Z$. This restricts the ${c^i_j}$ to the choice 
%\ref{choice}.

Here an important point should  be noticed:   the Lagrangian  $\cal L $ of
eq.~(\ref{lagrr})    is  reduced  to a simple function of a finite number of
$b_j^\alpha$ variables, by means of 
eq.~(\ref{source}) with the constraint given in eq.~(\ref{choice}).  
Namely, plugging 
the explicit expressions of eq.~(\ref{eqphi}) into   the Lagrangian
of eq.~ (\ref{lagrr}) we obtain                  
\ba    
{\cal{L}}&=&{1\over 2 } \int \; {d^4 \bar p}~ {d^4 \bar q }~
\bar \delta(p+q) ~              
\tilde\Pi^{\alpha\beta}(P_j^2)~ \bar \delta(p-P_j)~
\bar \delta(q-P_r) \, b^\alpha_j~ b^\beta_r +\nn\\
&+&{1\over 6} \int{{d^4 \bar p}~ {d^4 \bar q}~
{d^4 \bar k}~
\bar\delta(p+q+k)~ 
 {\cal{O}}^{\alpha\beta\gamma}(p,q,k)}~\bar\delta(p-P_j)~
 \bar\delta(q-P_r)~  \bar\delta(k-P_t) ~ b^\alpha_j ~b^\beta_r~ b^\gamma_t \\\nn
\ea 
We define the matrices
\be  
\Delta_{jl}^{\alpha\beta}=\int{{d^4 \bar p}~ {d^4 \bar q}~
 \bar\delta(p+q)~ 
\tilde\Pi^{\alpha\beta}(P_j^2)~  \bar\delta(p-P_j)~
 \bar\delta(q-P_l) } 
\ee 
and 
\be
D_{jlm}^{\alpha\beta\gamma}  =\int{{d^4 \bar p}~ {d^4 \bar q}~
{d^4 \bar k}~
\bar \delta(p+q+k)~ 
 {\cal{O}}^{\alpha\beta\gamma}(p,q,k)}~\bar\delta(p-P_j)~
 \bar\delta(q-P_l) ~ \bar\delta(k-P_m). 
\ee
After integration they become: 
\ba
D_{jlm}^{\alpha\beta\gamma}&=&
   \left\{ \matrix{
     {\cal{O}}^{\alpha\beta\gamma}(P_j,P_l,P_m) & \mbox{if} \quad 
                                      P_j+P_l+P_m=0 \cr\cr
      0 & \mbox{if} \quad 
                                      P_j+P_l+P_m \ne 0 } \right.
           \\                                          
\Delta_{jl}^{\alpha\beta}&=&
      \left\{ \matrix{
      \tilde\Pi^{\alpha\beta}(P_j^2) & \mbox{if} \quad
                                   P_j+P_l=0 \cr \cr
      0 & \mbox{if} \quad
                                   P_j+P_l \ne 0. } \right.
\ea
We recall that, in the above expressions, 
$ {\cal{O}}^{\alpha\beta\gamma}$ is  the interaction 
in momentum space,   $ \tilde\Pi^{\alpha\beta}$ is the inverse propagator 
and the quantities
 $ D_{jlm}$ and $\Delta_{jl}$ satisfy the four-momentum conservation by 
construction.

The function $Z$  then becomes 
\be \label{eqz}        
Z
 =a_i^\alpha b_i^\alpha- \frac{1}{2}
  \, b_j^\alpha \, b_l^\beta  \, \Delta_{jl}^{\alpha\beta} \,-\, \frac{1}{6}
  \, b_j^\alpha \, b_l^\beta \, b_m^\gamma \, D_{jlm}^{\alpha\beta\gamma}.     
\ee                                           
Equation (\ref{eqmoto}) provides us with a set of  iterative relations 
  for  the $b_j^\alpha$: after the substitution in eq.~(\ref{eqphi}),
and performing 
the integrals over the momenta, we are left with a relation which gives us 
each  $b_j$ at the order $t$ ($b_{j,t}$) in the interaction coefficient $\cal
O$, in terms of the   $b_{j,r}$ (with $r<t$)\footnote{In the following
$j,k,l,m$ will always label quantities that are in one to one correspondence
with the momenta $P_j$ in eq.~(\ref{eqphi}), whereas $r,s,t$ will  denote the
order of the perturbative expansion.}
                                     
Namely, we have \cite{alpha} 
\ba
b_{j,0}^\alpha&=&a_j^\alpha \quad j=1,n \nonumber \\
 b_{j,0}^\alpha & = & 0 \quad j>n      
  \quad \quad \mbox{$n$= number of external particles} \nonumber \\
b_{j,1}^\alpha&=&-{1\over 2}(\Delta^{-1})_{j,m}^{\alpha \beta}
 \, {D}_{m,k,l}^{\beta\gamma\delta}\; \sum_{l\neq k} 
 \, b_{k,0}^\gamma b_{l,0}^\delta \, \label{algorithm} \\
  \vdots & & \vdots \nonumber \\                       
b_{j,t}^\alpha&=&-{1\over 2}(\Delta^{-1})_{j,m}^{\alpha \beta}
 \, {D}_{m,k,l}^{\beta\gamma\delta} \; \sum_{r+s=t-1}
 \, b_{k,r}^\gamma \, b_{l,s}^\delta \label{itstep} \nonumber 
\ea
where the condition $l \ne k$ derives from the constraint in 
eq.~(\ref{choice}).
The full scattering amplitude is recovered by plugging
$b_j^\alpha=\sum_{t} b_{j,t} $ in  eq.~(\ref{eqz})       
and keeping only terms which are proportional to 
$a_1^\alpha a_2^\beta \cdots a_n^\gamma$ (the only ones that contribute to the 
limit $a\rightarrow 0$)
\ba
{\cal A}_{p_1,...,p_n}&=&-{1\over 2}\sum_{s+r=n-2}b_{j,r}^\alpha~
\Delta^{\alpha\beta}_{j,l}b_{l,s}^\beta 
-{1\over 6}
\sum_{s+r+t=n-3}{D}_{j,k,l}^{\beta\gamma\delta}~ b_{j,r}^\beta~
b_{k,s}^\gamma~ b_{l,t}^\delta +
b_{j,n-2}^\alpha~ a_l^\beta ~\tilde\Pi^{\alpha\beta}_{j,l}.
\label{ampl}
\ea
Notice that in eqs.~(\ref{eqz})--(\ref{ampl}) repeated indices are summed over,
in particular repeated latin indices are summed from 1 to $2^n-2$, namely over
the non-zero momenta contributing to eq.~(\ref{eqphi}).
                                         
Equation (\ref{ampl}) can be further simplified using the equations of motion.
The solution $b_j$ to $\partial Z/ \partial b_j=0$, has been found for any
value  of the expansion  parameter, thus the  minimization is satisfied order
by  order in the  perturbative expansion, {\it i.e.}  
$\partial Z/ \partial b_{j,t}=0$.
  For simplicity, let us discuss the case of an odd
number of particles $n$. It is easy to check by inspection
that each term in eq.~(\ref{ampl}) can contain at most
one $b_{j,t}$ with $t>n/2-1$. Therefore we can reorganize eq.~(\ref{ampl})
by collecting  each $b_{j,t}$ with $t>n/2-1$. The coefficients of 
these terms  cannot depend on
other $b_{j,t^\prime}$ with $t^\prime >n/2-1$ (they are only 
polynomial in $b_{j,\bar t }$ 
with $\bar t<n/2-1$) and therefore we can simply drop such terms
from eq.~(\ref{ampl}). 
 The above simplification also applies to the case of
even              
$n$, and it   amounts to keeping only the trilinear terms in 
eq.~(\ref{ampl}) and                 
limiting the summation to a subset of the $b_j^\alpha$ \cite{alpha}.
Then, an additional restriction on the  quantities $c^i_j$ in 
eq.~(\ref{choice}) is obtained:
\be
{\cal A}_{p_1,...,p_n} \=
-{1\over 6}\sum_{j,k,l\in {\cal P}}\sum_{s+r+t=n-3} 
{\cal O}_{j,k,l}^{\beta\gamma\delta}~ b_{j,r}^\beta~
b_{k,s}^\gamma~ b_{l,t}^\delta  \; ,
\label{ampl2}                             
\ee                                 
with:
\be
j\in {\cal P} \quad \mathrm{if } \quad
\left\{                               
 \matrix{ & \sum_i c^i_j<n/2  \cr
    & \mathrm{or} \cr
   & \sum_i c^i_j=n/2, \quad \mbox{with } \; c^1_j=1 \, . } \right.
\ee                                                        

A final remark is in order here. In eqs.~(\ref{algorithm}) one has to take
properly into account Bose/Fermi statistics. Formally this can
be achieved by introducing a set of Grassman variables $\epsilon_j$,
so that $\epsilon_j\epsilon_k  + \epsilon_k \epsilon_j = 0$, 
and setting $ u(p) \rightarrow \epsilon u(p)$ for the fermion sources,
where  $u(p)$ is a vector of ordinary numbers. In practice this
means that 
each term in the sums of eqs.~(\ref{algorithm}) enters with a relative sign,
depending on the order of the  $b_j$. 
                                       
% Finally one should keep in mind that a simplification closely parallel
%to the one obtained in eq.~(\ref{simple}) do always occur. 

The advantage of the iterative eqs.~(\ref{algorithm}) is that  they can now be
easily  implemented  in a {\tt Fortran} code. 
Note that the recursive relations  have
been obtained for  a completely generic  Lagrangian, without using any
specific property of the interaction  ({\it e.g.} identities  on the
structure constants of SU(3), or the Lorentz  structure of the interaction
etc.). With this algorithm we can thus compute the  scattering amplitude for
any  physical initial and final states. In particular, any colour structure  
can be assigned to the external legs (while for example
the Berends-Giele's recursive relations~\cite{Berends89}
have only been derived for colour-ordered amplitudes) and weak bosons (as
well as other particles beyond the Standard Model) can be incorporated.
Finally, the algorithm  has an exponential  growth of the CPU time with the
number of external particles (instead of a factorial growth) and it can take
into account the particle masses without increasing the computing time.
                                                                       
As a final remark, we point out that 
the algorithm presented above can be further optimised in the specific case
of the QCD Lagrangian. In fact the pure Yang-Mills Lagrangian
\begin{equation}                
{\cal L}_{YM}=-\frac{1}{4} F_{\mu\nu}^a F_{\mu\nu}^a
\label{ym} 
\end{equation}     
can be rewritten as
\begin{equation}
{\cal L}_{YM}=-\frac{1}{2}B_{\mu\nu}^aB_{\mu\nu}^a + \frac{1}{4}
 B_{\mu\nu}^aF_{\mu\nu}^a
\end{equation} 
This form of the Lagrangian has the virtue that a single interaction
term of the form $B A A$ is left, instead of the two interaction terms
which are present in eq.~(\ref{ym}). This saves CPU time when performing
the iteration step given in eq.~(\ref{algorithm}).       
A further reduction  in the algorithm complexity
can be obtained by using the Coulomb gauge $A^a_0=0$, which reduces the number
of components of the field $A^a_{\mu}$ from four to three and those of    
the field $B_{\mu\nu}^a$ from six to three.                         

\section{Summing over colours}
The proper bookkeeping of the colour structure of QCD processes has been one of
the main ingredients in the simplification of the calculations of
multi-parton processes in QCD which took place in the past
years. It has been shown~\cite{Lathuile} that by expanding the matrix element
for a given QCD process in a particular colour basis, the coefficients of each
colour structure in this basis enjoy important properties which make their
calculation much simpler. As an example, the scattering amplitude for $n$
gluons with momenta $p_i^{\mu}$, helicities $\epsilon_i^{\mu}$ and
colours $a_i$ (with $i=1,\dots,n$), can be written
as:                                          
\be      \label{eq:dualg}
  M(\{p_i\},\{\epsilon_i\},\{a_i\}) \; = \;
  \sum_{P(2,3,\dots,n)} \; \mbox{tr}(\lambda^{a_1}\,\lambda^{a_2}\dots
       \lambda^{a_n}) \; A(1,2,\dots ,n) \; .
\ee                                          
The sum extends over all permutations $P$ of $(2,3,\dots ,n)$, and the
functions $A(1,2,\dots, n) $ (known as {\em dual} or {\em
colour-ordered} amplitudes) are gauge-invariant,
cyclically-symmetric functions of the gluons'
momenta and helicities\footnote{For simplicity we will just use the indices
$i=1,\dots,n$, as opposed to using the full symbols $p_i$ and $\epsilon_i$,
to specify the relevant permutation of momenta and
helicities.}. All of the colour-dependence is absorbed in the trace
coefficients, and the dual amplitudes are colour-independent. 
In these special bases some particular helicity              
amplitudes~\cite{Parke86} have very simple analitical
expressions~\cite{Lathuile,Berends88}, regardless of the number of external
partons, and all amplitudes obey recursive relations~\cite{Berends88} which
make it possible to
numerically evaluate them systematically for arbitrarily complex
processes~\cite{Kleiss89,Berends89,Berends90}.
Thanks to additional properties of these functions, only 
$(n-2)!$ of them are truly independent, and the remaining ones 
can be obtained from specific {\em linear} combinations~\cite{Berends90}.
                                                                        
Furthermore, when summing over colours the amplitude squared, different
orderings of dual amplitudes are orthogonal at the leading order in $1/N^2$:
\be      \label{eq:dualgsq}
  \sum_{{\rm col's}} \vert M(\{p_i\},\{\epsilon_i\},\{a_i\}) \vert^2 \; = \;
  N^{n-2}(N^2-1)   \sum_{P(2,3,\dots,n)} 
    \left[  \vert A(1,2,\dots ,n)   \vert^2  +\frac{1}{N^2}(\mbox{interf.})
    \right]\, .
\ee                                  
It is therefore possible to achieve an accuracy to leading-order in $1/N^2$ by
neglecting the evaluation of the subleading interferences, reducing
significantly the complexity of the numerical evaluations.

Similar expansions hold for processes involving one~\cite{uppsala,Berends88}
or more~\cite{Mangano88} quark pairs.                                       
In the case of amplitudes with one quark pair, for example, 
one has the following expansion:          
\be     \label{eq:dualq}
  M(q_\alpha,\bar{q}_\beta,\{p_i\},\{\epsilon_i\},\{a_i\}) \; = \;
  \sum_{P(1,2,\dots,n)} \; (\lambda^{a_1}\,\lambda^{a_2}\dots
       \lambda^{a_n})_{\alpha\beta} \; A(q,\bar{q},1,2,\dots ,n) 
\ee                                                     
where the gauge-invariant functions $A(q,\bar{q},1,2,\dots ,n)$ obey 
remarkable properties
similar to those of the gluonic amplitudes. 
                        
Dual amplitudes can be easily evaluated using the {\tt ALPHA} algorithm.
Since the dual amplitudes $A$ are independent of the number of colours, 
they can be calculated exactly by taking $N$ sufficiently large. 
Considering for example the case of an $n$-gluon
amplitude, we can choose $N > n$ and select the following set of $\lambda$
matrices to represent the  gluon colours $a_1,\dots,a_n$: 
\be     \label{eq:lambdajk}                             
\left( \lambda^{a_i} \right)_{jk} \; = \;  \frac{1}{\sqrt{2}}
  \delta_{i,j} \delta_{i+1,k} \quad (i=1,\dots,n-1) \quad
\quad                                                   
,
\quad
\left( \lambda^{a_n} \right)_{jk} \; = \;  \frac{1}{\sqrt{2}}
  \delta_{n,j} \delta_{k,1}                                   
\ee                                        
With this colour choice the dual amplitude corresponding to the permutation
$(1,2,\dots,n)$ is proportional to the full amplitude, as the only non-vanishing
colour factor in eq.~(\ref{eq:dualg}) is                           
\be                                   
  \mbox{tr}(\lambda^{a_1}\lambda^{a_2}\dots\lambda^{a_n}) \; = \;
   \frac{1}{2^{n/2}} \; .
\ee
We explicitly calculated multi-gluon dual amplitudes using the large-$N$
colour-matrices given in eq.~(\ref{eq:lambdajk}). We verified the correctness
of the calculation for $n$ up to 11
by comparing the results for maximally helicity violating (MHV)
amplitudes~\cite{Parke86} (e.g. $g^+ g^+ \to g^+ \cdots g^+$)
with the analytic expressions       
known exactly for arbitrary $n$~\cite{Lathuile,Berends88}. 
The input of the numerical evaluation of the matrix element is a string
containing the total number of gluons, their helicity state,
and their momenta. From these data, the amplitude is evaluated
automatically. Since the evaluation performed with the 
{\tt ALPHA} algorithm does not treat the case of MHV amplitudes differently
than any other helicity combination, and since we checked the numerical
calculation using several different gauges (which allowed us to ensure that no
accidental cancellation of individual diagrams takes place), we are confident
that our code correctly evaluates the dual amplitudes for arbitrary helicity
configurations. We extended this test to processes with one and two $q\bar{q}$
pairs, where analytic expressions are known for similar MHV
amplitudes~\cite{review}, and found perfect agreement.
                                                      
The use of dual amplitudes for the evaluation of multi-parton processes has one 
valuable feature, and one serious drawback. The valuable feature is the fact
that dual amplitudes admit a simple physical interpretation, which as we shall
show makes them the required starting point for the 
parton-shower evolution of the parton-level process.
The serious drawback is that their number grows factorially with $n$, so that
while each individual dual amplitude can be easily and quickly calculated,
using the technique given above, the number of dual-amplitude calculations
which are necessary to get the full matrix element squared and summed over
colours becomes soon too large to be practical.
In the rest of this Section we will first explain in more detail the role of
dual amplitudes for the parton-shower evolution of the hard process, and then
present a way to bypass the rapid growth of their multiplicity.

Dual amplitudes correspond to planar amplitudes in the $N\to \infty$ limit of
QCD. At large $N$, the colour structure given by the assignement in
eq.~(\ref{eq:lambdajk}) corresponds to having the colour flowing from gluon 1
to gluon 2, from gluon 2 to gluon 3, and so on, until the colour of the last
gluon flows back to gluon 1 (see fig.~1).  
\begin{figure}
\centerline{\epsfig{figure= 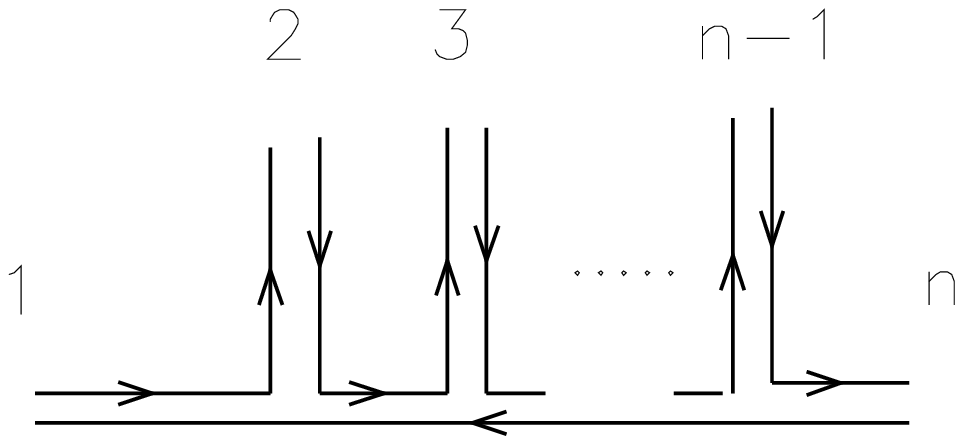, width=0.55\textwidth}}
\ccaption{}{Colour structure of the $n$-gluon amplitude in the large-$N$
limit.}
\end{figure}
The identification of a specific
colour flow makes it possible to incorporate soft-gluon emission corrections to
the hard process. In fact the soft-gluon emission probability 
from a planar amplitude is given, in the large-$N$ limit, by the incoherent
sum over the emission probabilities from each individual 
colour-string~\cite{bcm,Lathuile,Berends89}. In the case of a multi-gluon
amplitude, for example, we get:
\be                           
   \sum_{{\rm col's}} \vert M(p_1, \dots, p_n,k) \vert^2
   \; \buildrel{k \rightarrow 0}\over{\rightarrow} \;
   g^2 N
     \sum_{P(1,2,\dots,n)} 
     \sum_{i=1,\dots,n} \frac{(p_i \cdot p_{i+1})}{(p_i \cdot k)\, 
           (p_{i+1}\cdot k)}
  \; 
     \vert A(1,2,\dots, n) \vert^2 + {\cal
   O}(1/N^2) \; ,      
\ee              
where $n+1$ and $1$ are identified in the above equation. 
Inclusion of soft-gluon virtual corrections factorises in a similar fashion,
and Sudakov form factors for the soft-gluon emission probability can be
defined, to next-to-leading logarithmic accuracy, to describe the 
parton-shower evolution from any such colour-ordered 
process~\cite{bcm,herwig}.
Perfectly similar equations can be written in the case of processes involving
quark pairs.

The prescription to correctly generate the parton-shower associated to a given
event in the large-$N$ limit is therefore the following:
\begin{enumerate}
\item Calculate the $(n-1)!$ dual amplitudes corresponding to all possible
planar colour configurations.    
\item Extract the {\em most likely} colour configuration for this event    
on a statistical
basis, according to the relative contribution of the single configurations to
the total event weight~\footnote{Defining $w_i=\vert A_i \vert^2$ for each
colour flow $i$, and $W_i=\sum_{k=1,\dots,i} \, w_k/\sum_{k=1,\dots,n} \, w_k$,
the $j$-th colour structure will be selected if $W_{j-1} \le \xi < W_j$, for a
random number $\xi$ uniformly distributed over the interval $[0,1]$.}. Since
each dual amplitude is gauge invariant, the choice of colour-configurations is
also a gauge-invariant operation.
\item Develop the parton shower out of each initial and
final-state parton, starting from the selected colour configuration. This step
can be carried out by feeding the generated event to a Monte-Carlo program such
as {\tt HERWIG}, which is precisely designed to {\em turn partons into jets}
starting from an assigned colour-ordered configuration.
\end{enumerate}                                                        
Notice that, if the dual amplitudes are evaluated for a specific helicity
configuration, {\tt HERWIG} will also include spin-correlation effects in the
evolution of the parton shower~\cite{collins88,herwig}.
                                                   
With the physical value of $N=3$, dual amplitudes
corresponding to different permutations will however interfere with each other,
as shown in eq.~(\ref{eq:dualgsq}).                  
This interference is suppressed by powers of $1/N^2$~\cite{uppsala}, as well as
by dynamical factors: for example, the behaviour of interference terms is less
singular near collinear or soft momentum configurations than the leading terms
in $N$.
Within the $1/N^2$ approximation which is employed  in the description of
coherence effects in the shower evolution~\cite{herwig}, 
it is therefore consistent to
neglect the interferences between different dual amplitudes for the selection
of the colour structure to be assigned to a given event. After
calculating the total weight of a given event, accurate       
to all orders in $1/N$, we can thus still use the procedure
described in point 2 above to assign a definite colour configuration
to the event itself.

As a result, use of the dual-amplitude representation of a multi-gluon
amplitude allows to accurately describe not only the large-angle correlations
in multi-jet final states, but also the full shower evolution
of the initial- and final-state partons with the same accuracy available in
{\tt HERWIG} for the description of 2-jet final states. 
                                                  
In the $N\to \infty$ limit the choice of the dual amplitude basis has also one
important advantage. Since interferences between different colour structures
vanish in this limit, one can perform the sum over colours by Monte-Carlo
methods. Rather than evaluating the full matrix element squared, summed over
all colour structures, one can randomly select a dual colour structure on an
event-by-event basis, and just evaluate the corresponding contribution to the
amplitude squared. An overall multiplicative coefficient proportional to the
number of dual colour configurations provides the correct normalization.
Assuming that, on average, all colour configurations contribute the same amount
to the cross-section, this approach is numerically more efficient than summing
each event over all colours. Furthermore, one could optimise the
selection of colour configurations, to adapt it to possible differences in
their individual overall contributions.  This is similar to what is usually 
done to perform the sum over quark and gluon helicities.

At finite $N$ this procedure is not applicable anymore, as interferences
between various colour structures do not vanish. At the same time, 
the matrix describing all possible colour interferences has a size growing like
$[(n-1)!]^2$, which makes its storing and access highly inefficient.
            
We propose to solve this problem as follows. 
First of all we 
choose the following orthonormal basis for the Gell-Mann $\lambda$ matrices:
\ba                                          
&&   \lambda^1 \= \frac{1}{\sqrt{2}} \left( \begin{array}{ccc}
                          0 & 1 & 0 \\
                          0 & 0 & 0 \\
                          0 & 0 & 0 \end{array} \right) \quad,\quad
     \lambda^2 \= \frac{1}{\sqrt{2}} \left( \begin{array}{ccc}
                          0 & 0 & 1 \\
                          0 & 0 & 0 \\
                          0 & 0 & 0 \end{array} \right) \quad,\quad
     \lambda^3 \= \frac{1}{\sqrt{2}} \left( \begin{array}{ccc}
                          0 & 0 & 0 \\
                          1 & 0 & 0 \\
                          0 & 0 & 0 \end{array} \right) \nn \\
&&   \lambda^5 \= \frac{1}{\sqrt{2}} \left( \begin{array}{ccc}
                          0 & 0 & 0 \\
                          0 & 0 & 1 \\
                          0 & 0 & 0 \end{array} \right) \quad,\quad
     \lambda^6 \= \frac{1}{\sqrt{2}} \left( \begin{array}{ccc}
                          0 & 0 & 0 \\
                          0 & 0 & 0 \\
                          1 & 0 & 0 \end{array} \right) \quad,\quad
     \lambda^7 \= \frac{1}{\sqrt{2}} \left( \begin{array}{ccc}
                          0 & 0 & 0 \\    
                          0 & 0 & 0 \\
                          0 & 1 & 0 \end{array} \right) \nn \\
&&   \lambda^4 \= \frac{1}{{2}} \left( \begin{array}{ccc}
                          1 & 0 & 0 \\
                          0 &-1 & 0 \\
                          0 & 0 & 0 \end{array} \right) \quad,\quad
     \lambda^8 \= \frac{1}{\sqrt{12}} \left( \begin{array}{ccc}
                          1 & 0 & 0 \\    
                          0 & 1 & 0 \\
                          0 & 0 &-2 \end{array} \right) \nn 
\ea                                                       
In this basis, only a fraction of all possible $8^n$ colour assignements gives
rise to a non-zero amplitude. 
For each event, we randomly select a non-vanishing
colour assignement for the exernal gluons,
and evaluate the amplitude $M$. The weight of
the event is proportional to  $\vert M \vert^2$, multiplied by the
number of non-zero colour configurations. This is all we need
if we are not interested in evolving the event with the parton shower.
If instead we want to generate the parton shower, we first decide,
with standard unweighting techniques, whether to accept the event. 
If the event is accepted, we list all dual amplitudes               
contributing to the chosen colour configuration according to
eq.~(\ref{eq:dualg}) and, among these dual 
amplitudes, we randomly select a colour flow on the basis of their
relative weight\footnote{We explicitly checked the numerical
implementation of this algorithm by comparing our results for the colour-summed
squared amplitudes of the $gg \to {\rm n}\, g$ and 
$q \bar q \to {\rm n}\, g$ processes (for $n=4,5$) with the known results
obtained in ref.~\cite{Berends90}, as implemented in the {\tt NJETS}
code~\cite{Berends89b}. The agreement is at the level of machine precision.}.

In a 6-gluon amplitude, for example, 
a possible non-zero colour assignement is given by              
$(2,7,5,6,1,3)$.
Up to cyclic permutations, only three orderings of the colour indices give
rise to non-vanishing traces:
${\rm tr}(\lambda^2\,\lambda^7\,\lambda^5\,\lambda^6\,\lambda^1\,\lambda^3)$,        
${\rm tr}(\lambda^2\,\lambda^6\,\lambda^1\,\lambda^5\,\lambda^7\,\lambda^3)$ and 
${\rm tr}(\lambda^2\,\lambda^7\,\lambda^3\,\lambda^1\,\lambda^5\,\lambda^6)$. 
\begin{figure}                        
\centerline{\epsfig{figure= 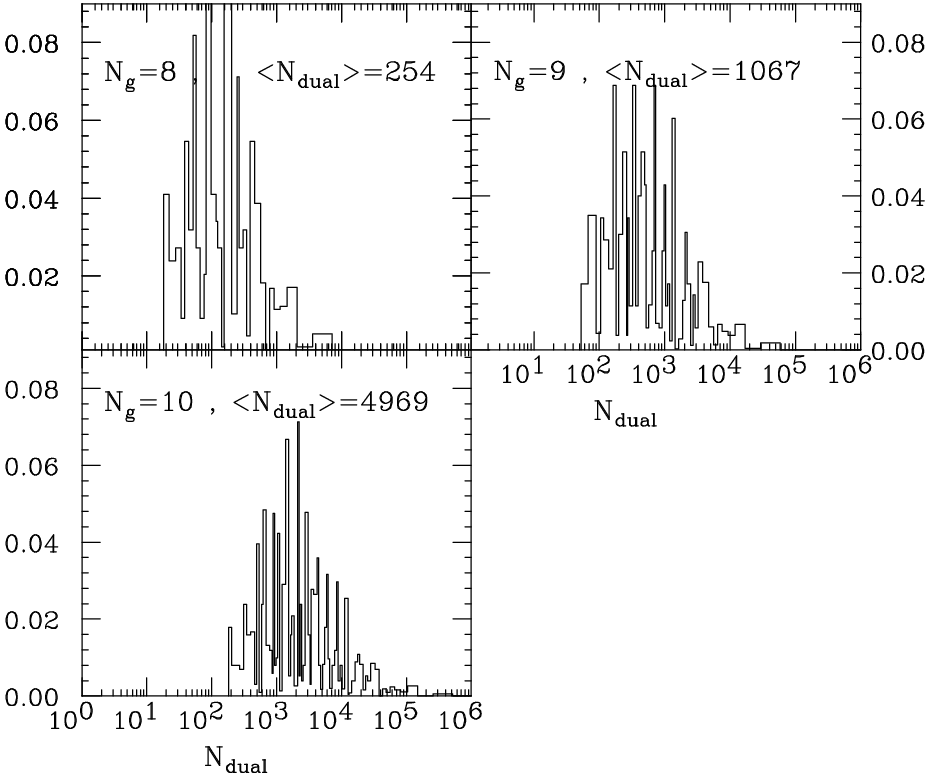,width=0.7\textwidth}}
\ccaption{}{Distribution of the number of dual amplitudes contributing to all
possible colour assignements in the orthogonal basis for $n=8,9$ and 10
gluons.}
\end{figure}
Therefore only three dual
amplitudes contribute to the full amplitude: $A(2,7,5,6,1,3)$,
$A(2,6,1,5,7,3)$ and $A(2,7,3,1,5,6)$. The colour ordering to be specified for
the coherent parton-shower evolution can be selected by comparing the size of
the squares of ${\rm tr}(\lambda^2\,\lambda^{i_2}\dots\lambda^{i_6}) \,
A(2,i_2,\dots,i_6)$ for the three contributing permutations $(i_2,\dots,i_6)$ 
of the colour indices.            
                                           
Since the average number of dual amplitudes involved in the evaluation
of a single element of the orthogonal basis is smaller than $(n-1)!$, 
the complexity of the procedure grows more slowly than for the    
calculation done using directly the dual basis. 
The distribution of the number of dual amplitudes contributing to all possible
colour assignements in the orthogonal basis
for $n=8,9$ and 10 gluons is shown in fig.~2.
Furthermore, considering that
only unweighted events are usually evolved by the parton shower, and that
unweighted events are a small fraction of all generated parton-level
events, the decomposition in terms of dual amplitudes only needs to be
performed for a small fraction of the generated configurations.
                                                               
The results we present in the following are all relative to parton-level
results, and therefore only the Monte-Carlo summation over orthogonal colour
configurations is considered.

\section{Results}
As an example of our technique, we present here results for the
following two parton-level processes:
\begin{eqnarray}                     
g~g      &\to& 8~g  \nonumber \\
q~\bar q &\to& 8~g\,.
\end{eqnarray}
For comparisons, we also computed the above reactions in the 
simple approximation first suggested by Kunszt and Stirling~\cite{sphel}. 
This approximation (hereafter referred to as {\tt SPHEL}) consists in assuming
that the average value of MHV amplitudes is equal to             
the average value of all other non-zero amplitudes. In the case of $n$-gluon
amplitudes this amounts to estimating the sum over all helicity configurations
using the relation:                   
\be
    \sum_{\rm{hel's}} \; \left | M\left( gg \to ({\rm n}-2) \, g \right)
    \right|^2\=                                                  
    \frac{2^n-2(n+1)}{n(n-1)} \;
    \sum_{\rm{MHV}} \; \left\vert M\left(gg \to ({\rm n}-2) \, g\right)\right|^2
\ee                                                                   
where the sum on the right-hand side runs over all MHV
amplitues (e.g. $++ \to +\dots +$, $+- \to - +\dots +$, etc.). Their value is
known exactly for all $n$ at the leading-order in $1/N$~\cite{Parke86}:
\ba                         
  \sum_{\rm{MHV}} \; \left\vert M\left(gg \to ({\rm n}-2)\, g\right)\right|^2 \=
    4 \left(\frac{g^2 N}{2}\right)^{n-2}\, (N^2-1) \,                 
    \sum_{i,j} \; (p_i \cdot p_j)^4 \; \nn \\
\label{eq:PT}                              
    \sum_{P(1,n-1)} \frac{1}{(p_1\cdot p_2)(p_2\cdot p_3) \cdots (p_n\cdot p_1)}
\ea                                                                        
where the sum runs over all permutations $P(1,\dots,n-1)$ of the indices
$(1,\dots,n-1)$.                                                     
Similarly, in the case of $q\bar q g\dots g$ amplitudes, the {\tt SPHEL}
approximation amounts to assuming:
\be                               
    \sum_{\rm{hel's}} \; \left\vert M\left(q \bar q \to {\rm n} \, g\right) 
    \right\vert ^2 \=         
    \frac{2^{n-1}-1}{n} \;                            
    \sum_{\rm{MHV}} \; \left\vert M\left(q \bar q \to  {\rm n} \, g\right) 
   \right\vert^2
\ee                                                          
where the sum on the right-hand side runs over all MHV
amplitues (e.g. $+- \to - +\dots +$), whose value is  
known exactly for all $n$ at the leading-order in $1/N$~\cite{uppsala}:
\ba                         
    \sum_{\rm{MHV}} \; \left\vert M\left(q \bar q \to {\rm n} \, g\right) 
    \right| \=
    4 \left(\frac{g^2 N}{2}\right)^{n}\, (N^2-1) \,
    \sum_{i=1,n} \; \left [ (p_q \cdot p_i)^3 (p_{\bar q} \cdot p_i) 
                          + (p_q \cdot p_i) (p_{\bar q} \cdot p_i)^3 \right ]
\; \nn \\                                                                 
\label{eq:MP}                                    
    \frac{1}{p_q \cdot p_{\bar q} }
    \sum_{P(1,\dots,n)}             
    \frac{1}{(p_q\cdot p_1)(p_1\cdot p_2) \cdots (p_n\cdot p_{\bar q})} \; ,
\ea                                                                         
where the sum runs over all permutations $P(1,\dots,n)$ of the indices
$(1,\dots,n)$.                                                     
                           
The kinematic configuration and the cut values used in our numerical examples
are as follows:
\begin{eqnarray}  \label{eq:cuts}
\sqrt{\hat s}= 1500\,{\rm GeV}\,, \quad 
 p_{T_i} > 60\,{\rm GeV}\,, \quad |\eta_i| < 2\,, \quad \Delta R_{ij} > 0.7\,.
\end{eqnarray}                                          
These values, and the choice of a fixed strong coupling $\alpha_s= 0.12$,
only serve for illustrative purposes. A more complete
phenomenological analysis of production cross-sections and a comparison of
exact and approximate expressions will be presented elsewhere.
                                            
The integration over the phase space allowed by the cuts was performed
by Monte Carlo using both {\tt RAMBO}~\cite{rambo}, a flat phase space
generator, and a multichannel approach~\cite{multi}.         
Both integration strategies gave compatible and comparable efficiencies.
The sum over helicity configurations was performed via a flat
Monte Carlo generation. No attempt has been made to optimise 
this step. 
Some of the details of the numerical performance of the algorithm are given at
the end of the Section.

In Table~2 we present our Monte Carlo estimate for the partonic 
cross-sections, together with the values obtained by using {\tt SPHEL}. 
{\tt SPHEL} overestimates by about a factor 2 the exact cross-section 
for the process $ g~g  \to 8~g$ , while it is accurate at
the 10\% level for $q~\bar q \to 8~g$.

%\begin{figure}[h]
\begin{table}  
\begin{center}
\begin{tabular}{|c|c|c|} \hline 
 \rule[-7 pt]{0 pt}{24 pt}
Process               & Exact (pb)     & {\tt SPHEL} (pb)\\
\hline                                                   
\hline
\rule[-7 pt]{0 pt}{24 pt}
$ g~g      \to 8~g$   & 0.719 $\pm$ 0.019  & 1.53 $\pm$ 0.03 \\ 
\hline
\rule[-7 pt]{0 pt}{24 pt}
$ q~\bar q \to 8~g$   & (\,0.901 $\pm$ 0.009\,)$ \times 10^{-3} $  
                      & (\,1.042 $\pm$ 0.008\,)$ \times 10^{-3} $ \\ 
\hline                                                       
\end{tabular}
\ccaption{}{Partonic cross-sections in pb at $\sqrt{\hat s} = 1500$ GeV,
with the cuts described in the text.}                                   
\end{center}
\end{table}

Notice the large ratio of the $gg$-initiated amplitude, relative to the $q\bar
q$ one. In the case of $2\to 2$ processes, this ratio  is of order 30 for a
reference scattering angle $\theta=\pi/2$, while here it is much larger. It is
easy to understand this result considering the structure of the MHV amplitudes
in the two cases. In the gluon-only case, eq.~(\ref{eq:PT}), the numerators
are dominated by the term $(p_1p_2)^4 \sim \hat s^4$, while in the 
$q\bar q$ case, eq.~(\ref{eq:MP}), only terms
proportional to $t$-channel momentum exchange appear. With the increase in the
number of final-state particles, the average momentum exchanged in the
$t$-channels becomes smaller, while $\hat s$ stays the same, and the ratio
$\hat s / \langle t \rangle^4$ becomes very large. 

\begin{figure}
\centerline{\epsfig{figure= 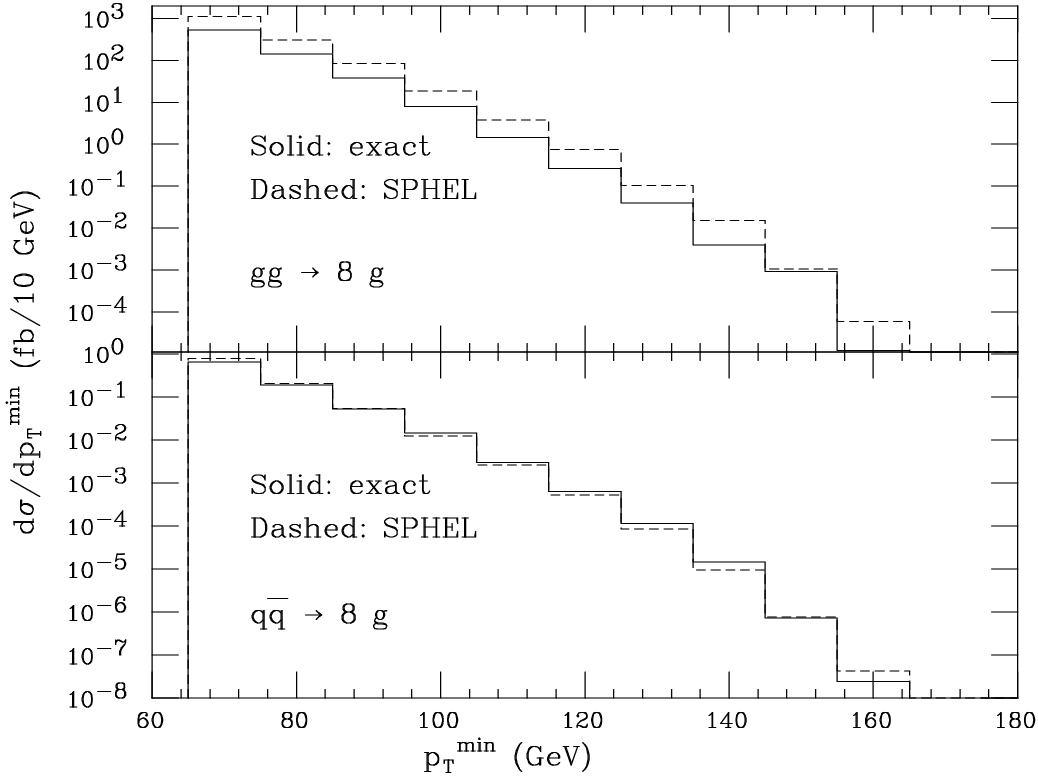, height= 9cm}}
\ccaption{}{Differential distributions for the minimum gluon 
transverse momentum. Exact result vs. {\tt SPHEL}.}
\end{figure}

\begin{figure}
\centerline{\epsfig{figure= 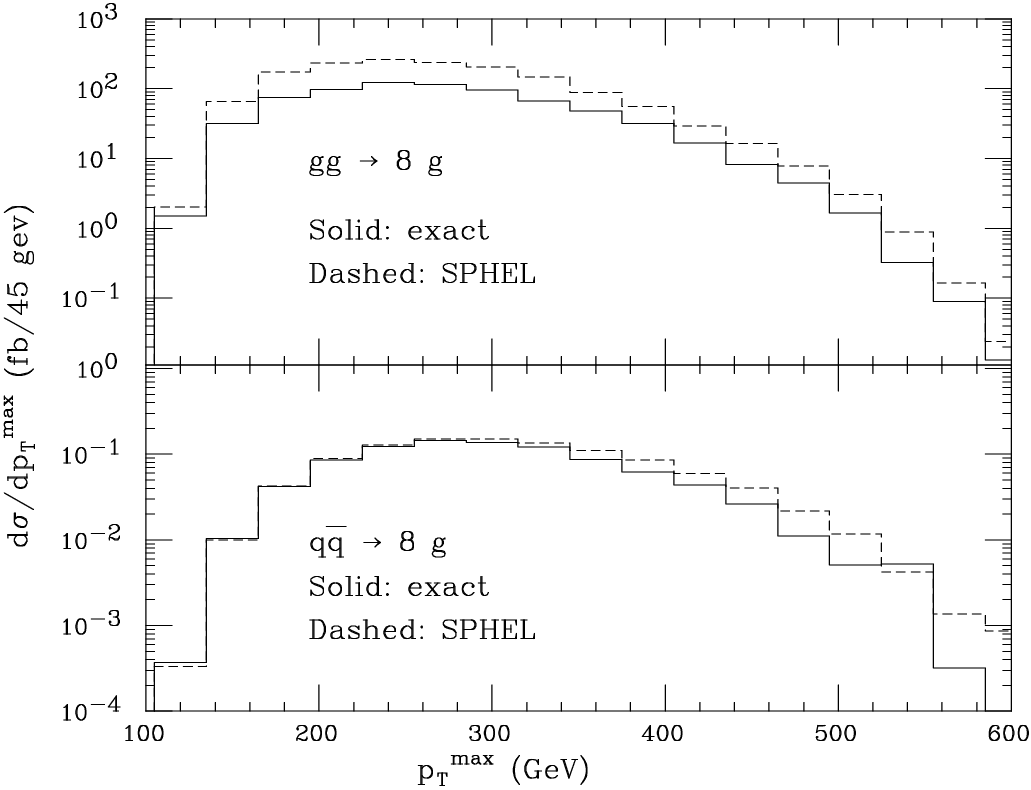, height= 9cm}}
\ccaption{}{Differential distributions for the maximun gluon 
transverse momentum. Exact result vs. {\tt SPHEL}.}
\end{figure}
\begin{figure}
\centerline{\epsfig{figure= 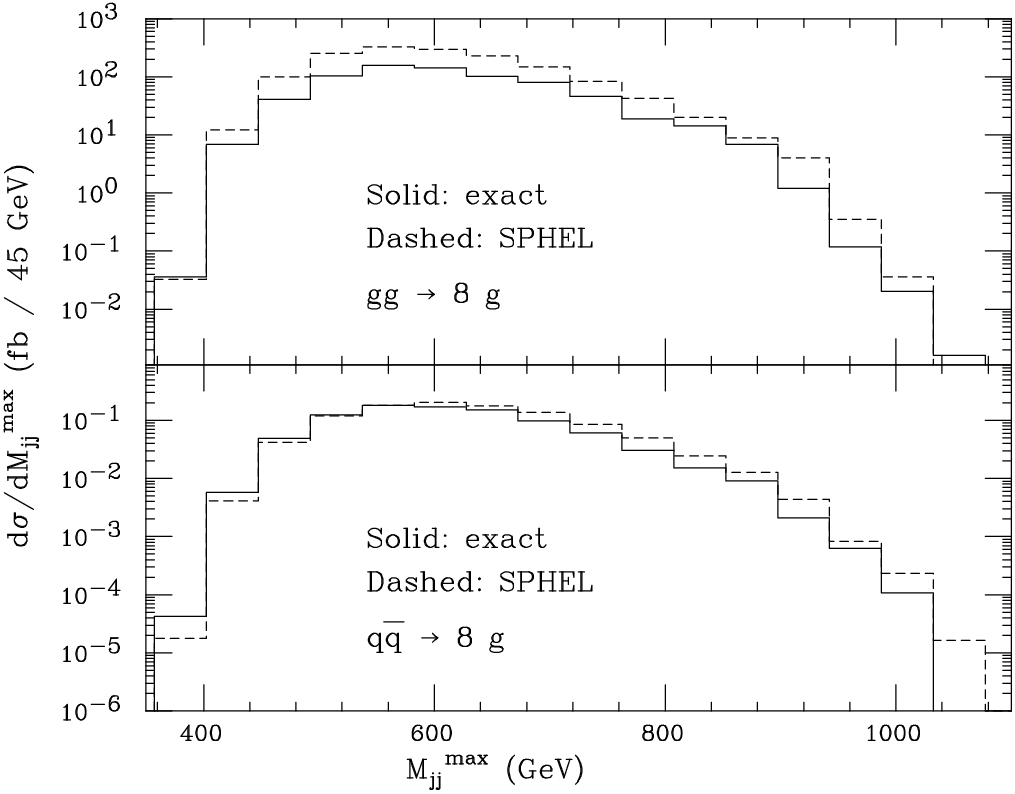, height= 9cm}}
\ccaption{}{Differential distributions for the maximum two-gluon 
invariant mass. Exact result vs. {\tt SPHEL}.}
\end{figure}

We also considered differential cross-sections. In Fig.~3, we show
the distribution of the minimum gluon transverse momentum for both
processes. A good agreement between the exact and the
{\tt SPHEL} result is observed. 
Considering the large uncertainties present in
the overall normalization of multi-parton tree-level processes (e.g. those
induced by changes in the choice of the renormalization and factorization
scales), the approximated evaluation can provide in many cases a sufficient
description of the final-state distributions.
                                             
In Figs. 4 and 5 we plot the distributions for the
maximum gluon transverse momentum and the maximum two-gluon invariant
mass. Again there is a reasonable agreement between the exact and the
approximated result, in particular in the case of the gluon-only process.
                                                                         
Before closing this Section, we present some technical details
to illustrate the performance of the algorithms used to produce
our results.                 
As an example, consider the process $g~g  \to 8~g$. 
As shown in Table~1, the total number of Feynman diagrams contributing to this
process is 10,525,900. The plots in this work have
been obtained from the evaluation of the matrix elements for $1.9 \times 10^6$
Monte Carlo points passing the selection cuts 
given in eq.~(\ref{eq:cuts}) (out of $1.8 \times 10^8$ points selected by the
phase-space generator). The efficiency for the generation of the non zero
weights, defined as the average weight divided by the maximum weight,
was about $1 \times 10^{-4}$. 
The computational time for producing 100 events that pass the cuts,
with a 200 MHz processor~\footnote{All of the following numbers are reduced by
a factor of 4 when using a DEC Alpha station.}, 
and working in double precision throughout,      
was about 91 seconds for the exact matrix                           
element and 2.3 seconds using the {\tt SPHEL} approximation. 
In this last case, the dominant component of the CPU budget is the search for
phase-space configurations passing the selection cuts.
                                                      
\section{Conclusions}
We presented in this work  an algorithm to evaluate the exact, tree-level
matrix elements  for multi-parton processes in QCD. This technique, based on
the algorithm {\tt ALPHA}, has been tested for processes such as $gg \to n$
gluons and $q\bar q \to n$ gluons, with $n$ up to 9. We discussed how the
summation over colour configurations  allows the construction of
parton-level event generators suitable to interfacing with a parton-shower
evolution including the effects of colour-coherence. This will eventually lead
to a fully exclusive, hadron-level description of multi-jet final states,
accurately incorporating the dynamics of large jet-jet separation angles.

While we confined our presentation to the case of hadroproduction, our
program can be used without modifications for the evaluation of multi-parton
production in $e^+e^-$ collisions, since the relevant $SU(2)\times U(1)$
Lagrangian is already included in the code. Likewise, associated
hadroproduction of gauge bosons and QCD partons is a straightforward
application of our code.
                                                                       
We gave some explicit numerical results, considering parton-level
cross-sections and distributions for the processes $gg \to 8$ gluons and $q\bar
q \to 8$ gluons. We verified that some standard simple approximations to the
multi-parton cross-sections provide a good description of the exact result, for
both rates and distributions. The existence of the exact results allows now to
extend the checks on these approximations to values of $n$ larger than ever
before.  We expect that large improvements can be obtained in the numerical
efficiency of the algorithm, and that cross-sections for up to 10 final-state
partons should become feasible. 

Furthermore, when $n$ becomes so large that the evaluation of the exact
amplitudes is numerically too slow for the generation of large samples of
events, one can nevertheless use the approximated calculations to generate the
samples, and lower statistics runs to assess the goodness of the approximation.
In this respect, the numerical efficiency of the phase-space generation in
presence of kinematic cuts will  need to be improved as well.
                                                             
\noindent {\bf Acknowledgements} We thank P. Nason for useful discussions.
The work of M. Moretti is funded by a Marie Curie fellowship (TMR-ERBFMBICT
971934).

\end{document}